\documentclass{PoS}
\usepackage{graphicx}
\usepackage[subfigure]{graphfig}
\usepackage{amsmath}
\usepackage{epsfig}
\usepackage{epstopdf}
\usepackage{color}
\def\be{\begin{equation}}
\def\ee{\end{equation}}
\def\bea{\begin{eqnarray}}
\def\eea{\end{eqnarray}}

\title{Quark Spin in Proton from Anomalous Ward Indentity
}

\ShortTitle{Quark Spin in Proton}

\author{Yi-Bo Yang$^{1}$, \speaker{Ming Gong}$^{2}$, Keh-Fei Liu$^{1}$, Mingyang Sun$^{1}$
\\
$^{1}$Department of Physics and Astronomy, University of Kentucky, Lexington, KY 40506\\
$^{2}$Institute of High Energy Physics, Chinese Academy of Sciences, Beijing 100049, China\\

\vspace*{-0.5cm}
\begin{center}
\large{
\vspace*{0.4cm}
\includegraphics[scale=0.20]{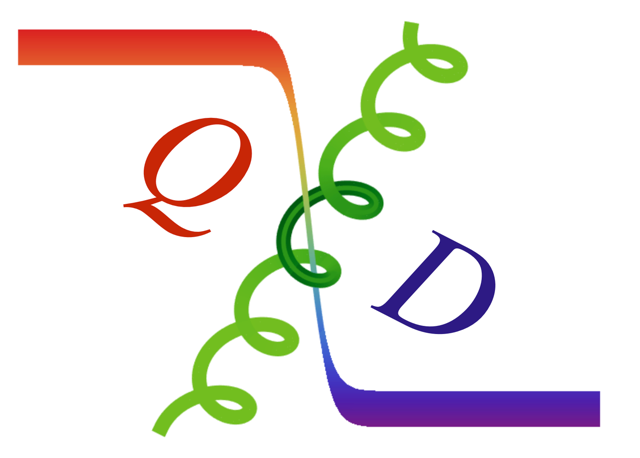}\\
\vspace*{0.4cm}
($\chi$QCD Collaboration)
}
\end{center}
}
\abstract{We report a quark spin calculation from the anomalous Ward identity with overlap fermions on $2+1$ flavor dynamical fermion configurations with light sea quark masses. Such a formulation decomposes the divergence of the
flavor-singlet axial-vector current into a quark pseudoscalar term and a triangle anomaly term, flavor by flavor. A large negative contribution from the anomaly term is observed and it is canceled within errors by the contribution from the pseudoscalar term in the disconnected insertion in the heavy quark region. On the other hand, net negative contributions are obtained for the light and strange quarks in the disconnected insertion, since their quark pseudoscalar terms  are smaller than that of the heavy quark. Our results are obtained from the $2+1$ flavor domain wall fermion configurations on the $24^3 \times 64$ lattice with $a^{-1} = 1.78(5)$ GeV and the light sea quark at $m_{\pi} = 330$ MeV. We use the overlap fermion for the valence and the quark loop so that the renormalization constants $Z_m$ and $Z_P$ cancel in the pseudoscalar operator $2 m P$. In addition, the overlap Dirac operator is used to calculate the local topological charge in the anomaly so that there is no renormalization for the anomaly term either. In this study, we find the total quark spin to be small mainlyly due to the large negative anomaly term which could be the source for the `proton spin crisis'.}

\FullConference{32st International Symposium on Lattice Field Theory - LATTICE 2014\\
        June 22 - June 29, 2014\\
        Columbia University\\
        New York, USA}

\begin{document}
\section{Introduction}
Apportioning the spin of the nucleon among its constituents of quarks and gluons is one of the most challenging issues in QCD both experimentally and theoretically.  It is shown by X. Ji~\cite{Ji:1996ek} that there is a gauge-invariant separation of the proton spin operator into
the quark spin, quark orbital angular momentum, and glue angular  momentum operators
\begin{equation}       
\vec J_{\rm{QCD}} = \vec J_q + \vec J_g
  = \frac{1}{2} \vec\Sigma_q + \vec{L}_q + \vec{J}_g ,
\label{ang_op_def_split_2}
\end{equation}
where the quark and glue angular momentum operators are defined from the
symmetric energy-momentum tensor
\begin{equation}
  J_{q,g}^i 
 = \frac{1}{2}\,\epsilon^{ijk}\,\int \, d^3x\, \left(\mathcal{T}_{q,g}^{0k}\, x^j
    - \mathcal{T}_{q,g}^{0j}\, x^k\right) ,
\label{ang_op_def_split_1}
\end{equation}
with the explicit expression
\begin{equation}
  \vec{J}_q = \frac{1}{2} \vec\Sigma_q + \vec{L}_q  =  \int d^3x \, 
  \bigg{[} \frac{1}{2}\, \overline\psi\,\vec{\gamma}\,\gamma^5 \,\psi 
 + \psi^\dag \,\{ \vec{x} \times (i \vec{D}) \} \,\psi \bigg{]} ,
\label{quark_ang_op_split_1}
\end{equation} 
for the quark angular momentum which is the sum of quark spin and orbital angular
momentum, and each of which is gauge invariant. 

Since the contribution from the quark spin is found to be small
($\sim$25\% of the total proton spin) from the global analysis of deep inelastic scattering 
data~\cite{deFlorian:2009vb}, it is expected that the remainder should come from glue spin and the orbital angular momenta of quarks and glue. The quark spin contribution from $u$,~$d$~and~$s$ has been studied on the 
lattice~\cite{Dong:1995rx,Fukugita:1994fh} since 1995 with quenched approximation or 
with heavy dynamical fermions~\cite{Gusken:1999as}.\ Recently,\ it has been 
carried out with light dynamical 
fermions~\cite{QCDSF:2011aa,Engelhardt:2012gd,Abdel-Rehim:2013wlz,Babich:2010at}
 for the strange quark.
 The strange quark spin ($\Delta s$) they found,  is in the range from $-0.02$ to $-0.03$
which is several times smaller than that from a global fit of DIS and semi-inclusive DIS (SIDIS)
which gives $\Delta s \approx - 0.11$~\cite{deFlorian:2009vb}. Such a discrepancy between the global fit of experiments and the lattice calculation
of the quark spin from the axial-vector current has raised a concern that the renormalization constant 
for the flavor-singlet axial-vector current could be substantially different from that of the 
isovector axial-vector current~\cite{Karsten:1980wd,Lagae:1994bv} since the latter is commonly use for
the lattice calculations of the flavor-singlet axial-vector current for the quark spin. To alleviate this concern, we  use the anomalous Ward identity (AWI) to calculate the quark spin. 
In this proceeding, we report the calculation of both the connected insertion (CI) and disconnected 
 insertion (DI) contributions to quark spin from $u$,\ $d$,\ $s$ and 
$c$, which are non-perturbatively renormalized when the anomalous Ward identity is used to
carry out the calculation with the overlap fermion for the quarks and the overlap Dirac operator 
for the topological charge.
   
The anomalous Ward identity includes the triangle anomaly in the
divergence of the flavor-singlet axial-vector current
\begin{equation}\label{awi}
\partial^{\mu} A_{\mu}^0 = 2 \sum_{f=1}^{N_f}  m_f \overline{q}_f
i \gamma_5 q_f + 2i N_f  q ,
\end{equation}
where $q$ is the local topological charge operator and is equal to 
$\frac{1}{16 \pi^2} tr_c G_{\mu\nu} \tilde{G}_{\mu\nu}$ in the continuum. We put this identity between
the nucleon states and calculate the matrix element on the right-hand side with a 
momentum transfer $\vec{q}$
\begin{eqnarray} \label{eq:awi}
\langle p^\prime s \left| A_\mu \right| p s \rangle s_\mu = \lim_{\vec{q} \rightarrow 0} \frac{i | \vec{s} |}
{\vec{q} \cdot \vec{s}}  \langle p^\prime,s  | 2 \sum_{f=1}^{N_f}  m_f \bar{q}_f i \gamma_5 q_f 
+ 2 i N_f q \, | p,s \rangle.
\end{eqnarray}
Lattice QCD theory has finally accommodated chiral symmetry, the lack of which has hampered the 
development of chiral fermions on the lattice for many years. It is shown that when the lattice massless Dirac operator satisfies the Gingparg-Wilson relation $\gamma_5 D + D \gamma_5 = a D\gamma_5D$,
for which the overlap fermion with negative mass parameter being the explicit example~\cite{Neuberger:1997fp}, the
modified chiral transformation leaves the action invariant and gives rise to a chiral Jacobian factor
$J = e^{-2i\alpha Tr \gamma_5 (1 - \frac{1}{2}a D)}$ from the fermion determinant~\cite{Luscher:1998pqa}.
The index theorem~\cite{Hasenfratz:1998ri} shows that this Jacobian factor carries the correct chiral anomaly.
It is shown further that the local version of the overlap Dirac operator gives the topological
charge density operator in the continuum~\cite{Kikukawa:1998pd}, i.e.
\begin{equation}  \label{top-charge}
Tr \gamma_5 (1 - \frac{1}{2}a D_{ov}(x,x) )= \frac{1}{16 \pi^2} tr_c \, G_{\mu\nu} 
\tilde{G}_{\mu\nu}(x) + \mathcal{O}(a^2)
\end{equation}
Therefore, Eq.~(\ref{awi}) is exact on the lattice for the overlap fermion which gives the correct 
anomalous Ward identity at the continuum limit. Instead of calculating the matrix element of
the axial-vector current derived from the Noether procedure~\cite{Kikukawa:1998py,Hasenfratz:1998ri},
we shall calculate it from the r.h.s. of the AWI in Eq.~(\ref{awi}) through the form factors defined
in Eq.~(\ref{eq:awi}). 

In the lattice calculation with the overlap fermion, we note that the renormalization constant 
of the pseudoscalar density cancels that of the renormalization of the quark mass, i.e. 
$Z_m\,Z_P = 1$ for the chiral fermion. Also, the topological charge density, when calculated 
with the overlap operator as in the l.h.s of Eq.~(\ref{top-charge}) is renormalized -- its integral
over the lattice volume is an integer satisfying the Atiya-Singer theorem.
Thus, when the matrix elements on the right-hand side of Eq.~(\ref{eq:awi}) are calculated with the overlap fermion and
its Dirac operator, the flavor-singlet axial-vector current is automatically renormalized on the lattice
non-perturbatively \mbox{\it{\`{a} la}} anomalous Ward identity (AWI). 

\section{Numerical details}

Besides the fact that AWI admits non-perturbative renormalization on the lattice, 
the pseudoscalar density and the topological density represent the low-frequency and
high-frequency parts of the axial-vector quark loop respectively. On the $2+1$ flavor
domain-wall fermion configurations on the
$24^3 \times 64$ lattice ($a^{-1}=1.78(5)$ GeV~\cite{Yang:2014sea}) with the light sea quark mass corresponding to a pion mass at 330 MeV. It is learned that 
on such a lattice, a mere 20 pairs of the overlap low eigenmodes would
saturate more than 90\% of the pseudoscalar loop with light quarks in configurations with 
zero modes~\cite{Gong:2013vja}. On the other hand, it is well-known that the
contribution to the triangle anomaly comes mainly from the cut-off part of the regulator.
Therefore, the topological charge density represents the high-frequency contribution
of the axial-vector loop, albeit in a local form (the overlap operator is exponentially
local). Since the quark loop of the pseudoscalar density is totally dominated by the low modes, we expect
that our low-mode averaging (LMA) approach should be adequate for this term. To the extent 
that the signal for the topological term is good, one should be able to calculate the flavor-singlet 
$g_A$ with the AWI . Both the overlap fermion for the quark loop and the overlap operator 
for the topological charge density are crucial in this approach.

 \begin{figure}[htb]
 \centering
  {\includegraphics[width=0.48\hsize,angle=0]{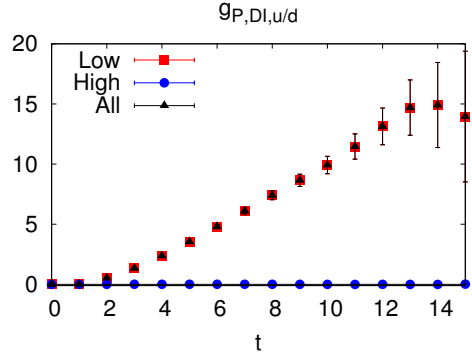}}
\caption{
The summed matrix elements for the light quark loop (corresponds to $m_{\pi} = 330$ MeV) with the pseudoscalr current, as a function of the separation of the source and sink. We confirm that the low mode part dominates in the pseudoscalar case.
 } \label{fig:high-low}
\end{figure}

We shall first check to see how the low mode averaging works for quark loops of the pseudoscalar operator. We plot the ratio of the 3-pt to 2-pt correlation functions in the DI for the sum method where the insertion is summed between the nucleon source and sink times so that the matrix element is the slope of the ratio at large source-sink time separation. This is done for the case where both the quarks in the loop and in the nucleon coincide with that of the light sea mass which corresponds to $m_{\pi} = 330$ MeV.  

 The reason for the good signal is that, besides adopting LMA in the DI, 
that low-mode substitution (LMS) technique with 8 smeared noise-grid source
on a given time slice~\cite{Gong:2013vja} has been used to improved the nucleon propagator in the construction
of the three-point function, and all the time slices has been looped over to improve the statistics.

\begin{figure}[hbt]
  \centering
\subfigure[]
 {{\includegraphics[width=0.48\hsize,angle=0]{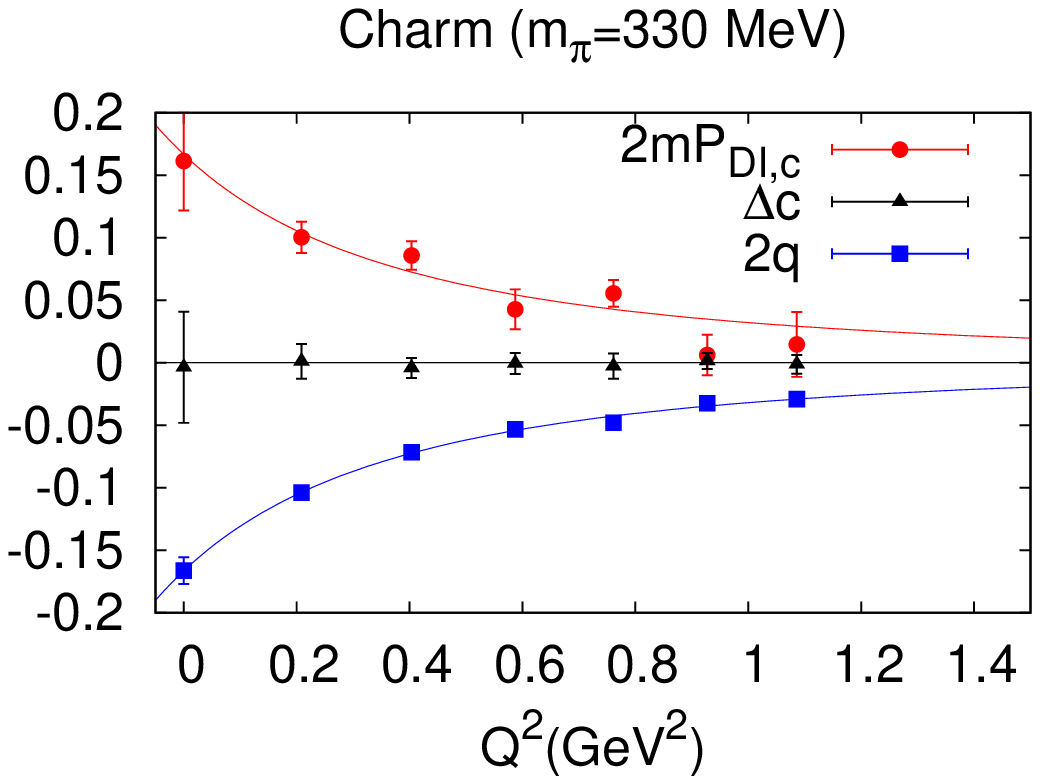}}
    \label{charm_spin}}
  \subfigure[]
  {{\includegraphics[width=0.48\hsize,angle=0]{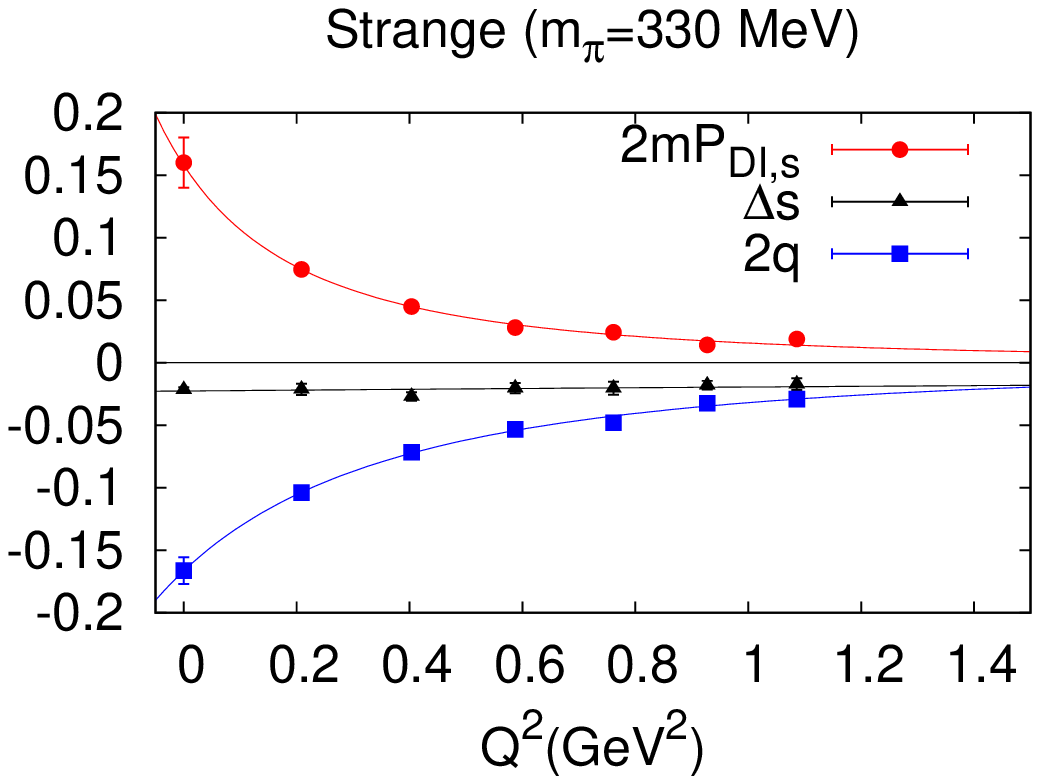}}
  \label{strange_spin}}  
  \caption{(a) The charm pseudoscalar and topological density contributions to the proton spin as
  a function of $Q^2$. (b) The same as in (a) for the strange.}
\end{figure}

We first show the results for the charm quark which contribute
only in the DI. The pseudoscalar density term and the topological charge density term
are plotted in Fig.~\ref{charm_spin} as a function of $Q^2$. We see that the
pseudoscalar is large due to the large charm mass and positive, while the topological 
charge term is large and negative. When they are added together (black triangles in the
figure), it is consistent with zero for the whole range of $Q^2$. When extrapolated to
$Q^2 =0$, the charm gives zero contribution to the proton spin within error due to the cancellation 
between the pseudoscalar term and the topological term. It is shown~\cite{Franz:2000ee} 
that the leading term in the heavy quark expansion of the quark loop of the pseudoscalar term, 
i.e. $2mP$ is the topological charge $\frac{2i}{16 \pi^2}tr_c G_{\mu\nu} \tilde{G}_{\mu\nu}$, 
but with a negative sign. Thus, one expects that there is no contribution to the quark spin from heavy quarks
to leading order. It  appears that the charm quark is heavy enough so that the $\mathcal{O}(1/m^2)$ correction
is small. We take this as a cross check of the validity of our numerical estimate
of the DI calculation of the quark loop as well as the anomaly contribution.

The contributions from the strange are also calculated and shown in Fig.~\ref{strange_spin}. The $2mP$ contribution
is slightly smaller than that of $2q$ and results in a net small negative value for the sum of $2mp$ and
$2q$ at finite $Q^2$. After a dipole fit, we obtain $\Delta s = -0.26(5)$ at $m_{\pi} = 330$ MeV.
 
 \begin{figure}[hbt]
  \centering
 \subfigure[] 
 {\includegraphics[width=0.48\hsize,angle=0]{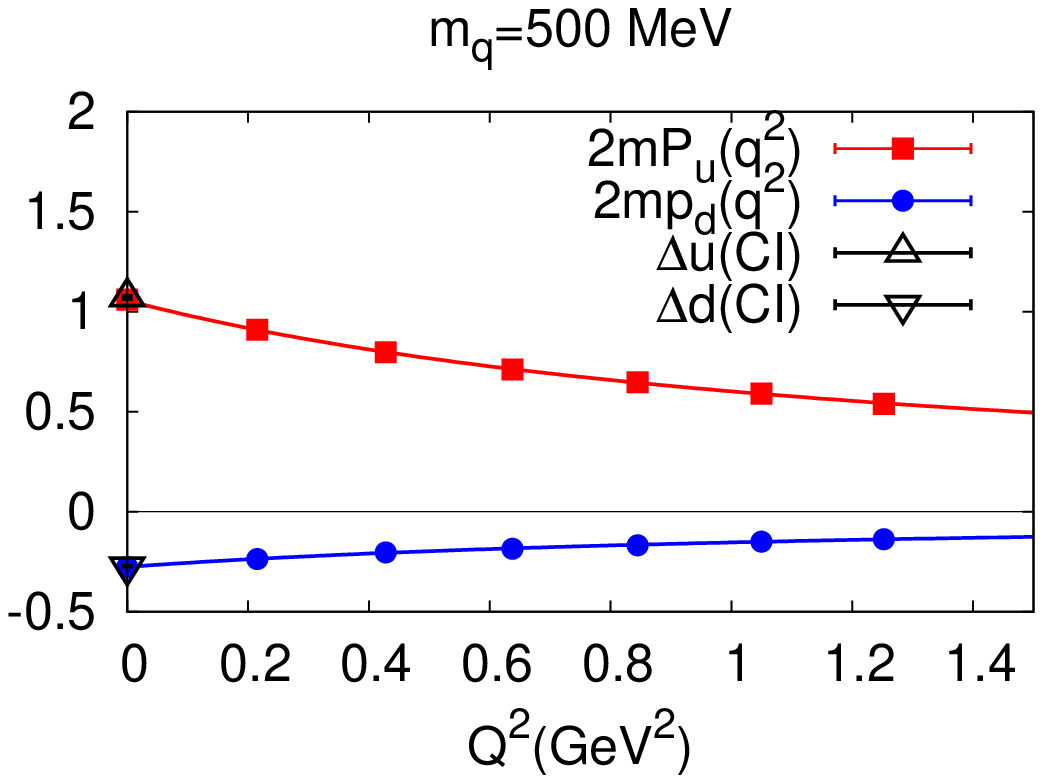}}
 \subfigure[]
 {\includegraphics[width=0.48\hsize,angle=0]{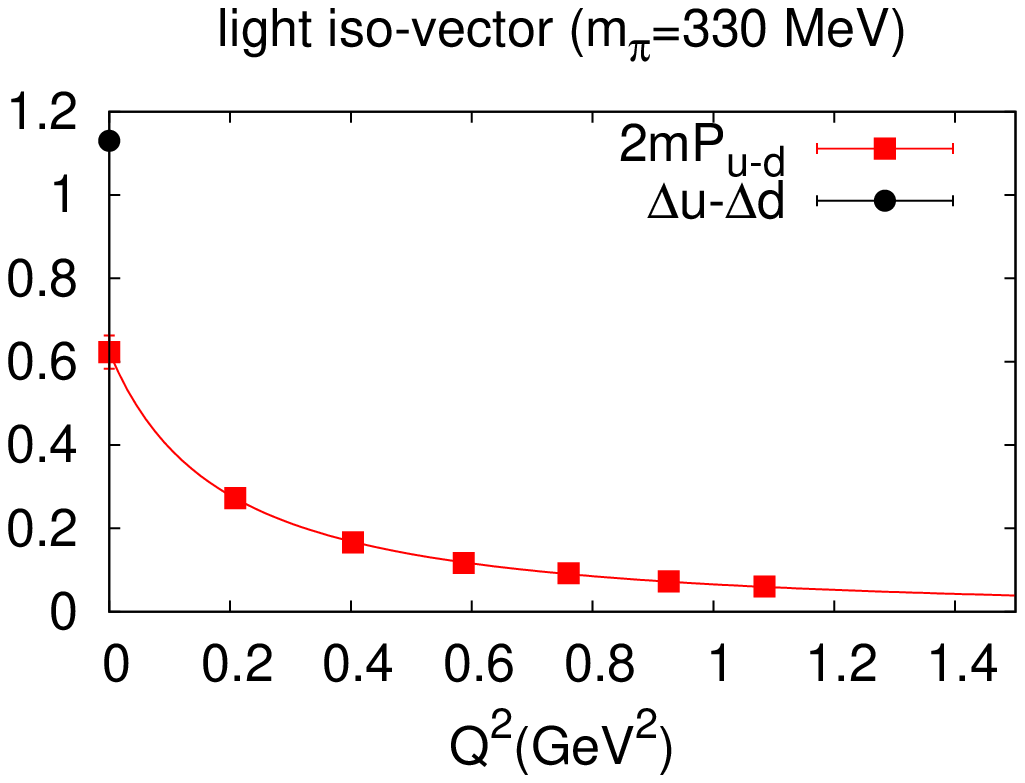}} 
  \caption{(a) The quark spin combinations of the proton-like baryon with $m_q\sim500$ MeV, in which the DI contribution is canceled by the topological change one. (b) The same as in (a) for light
quarks at the unitary point.}
  \label{fig:heavy}
\end{figure}

Since this $\Delta s$ is quite a bit smaller than the experimental value, we explore the possible finite volume effect and 
the fact that the induced pseudoscalar form factor $h_A(q^2)$ has been neglected in the $Q^2$ extrapolation which does not contribute at the $Q^2 = 0$ limit as in Eq.~(\ref{eq:awi}), but has a contribution at finite $Q^2$~\cite{Liu:1995kb}. We shall check this in the connected insertion (CI) calculation. As can be seen in Fig.~\ref{fig:heavy} for  $m_q\sim500$ MeV, both $\Delta u$ and $\Delta d$ in CI calculated from the axial-vector current and
renormalized with $Z_A$ from the isovector Ward identity are well reproduced through the $Q^2$ extrapolation  of
$2mP$ with a dipole form.  Whereas, in the case of light quarks at the unitary point, $g_A^3 = 1.13(2)$ from the axial-vector current is $1.8(1)$ times larger than $0.62(4)$ from the dipole extrapolation of $2mP$. This is most likely due to our ignoring of the induced pseudoscalar form factor $h_A(q^2)$ as well as the finite volume effect at small $Q^2$ which is well known to plague the extrapolation of nucleon magnetic form factor.

 \begin{figure}[htb]
 \centering
  {\includegraphics[width=0.7\hsize,angle=0]{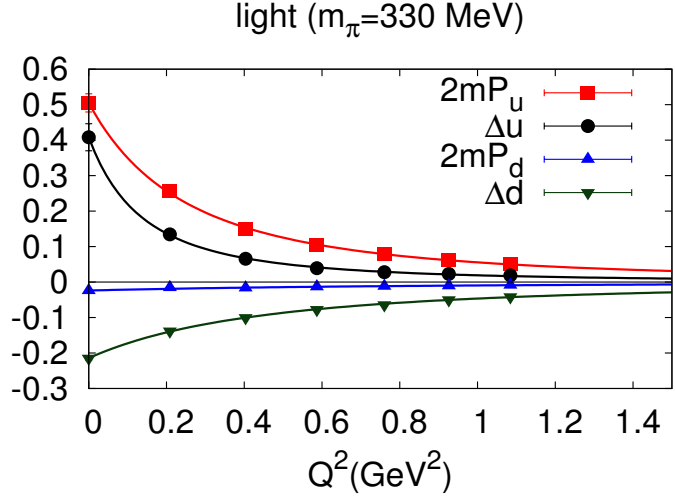}\label{ud_spin}
\caption{The combined pseudoscalar contribution from both the connected insertion (CI) and DI ($2mP_{u/d}$ in the plot), along with the overall quark spin from both pseudoscalar and topological charge ($g_{A,d}$). The plot corresponds to the unitary point with $m_{\pi} = 330$ MeV.}}
\end{figure}

At the unitary point, when the valence $u/d$ mass matches with that of the light sea, Fig.~\ref{ud_spin} shows the combined pseudoscalar contributions $2mP_{u/d}$  from both the combined CI and DI with a dipole extrapolation. Also plotted are the overall quark spin $\Delta u/\Delta d$ by including the topological charge contribution.  In this case, we obtain $\Delta u + \Delta d = 0.19(3)$ and $\Delta u - \Delta d = 0.62(4)$ at $Q^2 = 0$ from a dipole extrapolation in $Q^2$. As we discussed above, the fact that $g_A^3$ from the axial current is
$1.8(1)$ times larger than that of $\Delta u - \Delta d$ through the Ward identity approach is most likely due to the neglect of the induced pseudoscalar form factor $h_A(q^2)$ and the finite volume effect in the $Q^2$ extrapolation. We apply this  $1.8(1)$ factor as an estimate to correct the present AWI approach and obtain 
$\Delta u + \Delta d = 0.35(6)$, $\Delta s = -0.05(1)$. Thus the total estimated spin $\Delta \Sigma= 0.30(6)$ at the unitary point is consistent with the present experimental results which are between 0.2 and 0.3. We expect that, at lighter quark masses, $\Delta \Sigma$ will be smaller.
    
The above results are from the $24^3 \times 64$ lattice with 200 configurations.
The nucleon propagator in the DI has been calculated with the smeared-grid noise
source with time dilution which covers all time slices in order to have reasonable
statistics for the DI. 


\section{Summary}\label{sec:summary}

     We have carried out a quark spin calculation from the anomalous Ward identity with overlap fermions on $2+1$ flavor dynamical fermion configurations with light sea quark masses. A large negative contribution from the anomaly term is observed and it is canceled within errors by the contribution from the pseudoscalar term in the heavy quark region. On the other hand, net negative contributions are obtained for the light and strange quarks in the disconnected insertion, since their quark pseudoscalar terms  are smaller than that of the heavy quark. Since the overlap fermion is used for the
 pseudoscalar term $2mP$ and the overlap Dirac operator is used for the local topological term, the result from AWI is
 non-perturbatively renormalized.  Since in the CI, the dipole extrapolation of $2mP$ obtained a $\Delta u - \Delta d$
 which is smaller than that obtained from axial vector current with renormalization $Z_A$ from the Ward identity,
 one is concerned that there is substantial finite volume effect and also the fact that the induced pseudoscalar form factor
 $h_A(q^2)$ is neglected in the $Q^2$ extrapolation. These effects need to be studied  in the upcoming calculation on the $48^4 \times 96$ lattice which has twice the spatial size (i.e. 5.5 fm) of  the present one and the pion mass is at the physical value.
 
       As far as the smallness of the quark spin in the proton is concerned, present results from only one lattice notwithstanding, has presented hints to suggest that the culprit of the `proton spin crisis' is the $U(1)$ triangle anomaly.


\begin{thebibliography}{99}
\bibitem{deFlorian:2009vb} 
  D.~de Florian, R.~Sassot, M.~Stratmann and W.~Vogelsang,
  Phys.\ Rev.\ D {\bf 80}, 034030 (2009)
  [arXiv:0904.3821 [hep-ph]].
  %
\bibitem{Dong:1995rx}
  S.~J.~Dong, J.~-F.~Lagae, K.~F.~Liu,
  Phys.\ Rev.\ Lett.\  {\bf 75}, 2096-2099 (1995), 
  [hep-ph/9502334].
  %
  %
\bibitem{Fukugita:1994fh} 
  M.~Fukugita, Y.~Kuramashi, M.~Okawa and A.~Ukawa,
  Phys.\ Rev.\ Lett.\  {\bf 75}, 2092 (1995), 
  [hep-lat/9501010].
  %
  %
\bibitem{Gusken:1999as} 
  S.~Gusken {\it et al.}  [TXL Collaboration],
  Phys.\ Rev.\ D {\bf 59}, 114502 (1999).
  %
  %
\bibitem{QCDSF:2011aa} 
  G.~S.~Bali {\it et al.}  [QCDSF Collaboration],
  Phys.\ Rev.\ Lett.\  {\bf 108}, 222001 (2012), 
  [arXiv:1112.3354 [hep-lat]].
  %
  %
  \bibitem{Engelhardt:2012gd} 
  M.~Engelhardt,
  Phys.\ Rev.\ D {\bf 86}, 114510 (2012)
  [arXiv:1210.0025 [hep-lat]].
   %
   %
\bibitem{Abdel-Rehim:2013wlz} 
  A.~Abdel-Rehim, C.~Alexandrou, M.~Constantinou, V.~Drach, K.~Hadjiyiannakou, K.~Jansen, G.~Koutsou and A.~Vaquero,
  arXiv:1310.6339 [hep-lat].
  %
  %
\bibitem{Babich:2010at} 
  R.~Babich, R.~C.~Brower, M.~A.~Clark, G.~T.~Fleming, J.~C.~Osborn, C.~Rebbi and D.~Schaich,
  Phys.\ Rev.\ D {\bf 85}, 054510 (2012)
  [arXiv:1012.0562 [hep-lat]].
  %
    %
    \bibitem{Ji:1996ek} 
  X.~D.~Ji,
  Phys.\ Rev.\ Lett.\  {\bf 78}, 610 (1997)
  [hep-ph/9603249].
  %
  \%
 \bibitem{Karsten:1980wd} 
  L.~H.~Karsten and J.~Smit,
  Nucl.\ Phys.\ B {\bf 183}, 103 (1981).
 %
 %
  \bibitem{Lagae:1994bv} 
  J.~F.~Lagae and K.~F.~Liu,
  Phys.\ Rev.\ D {\bf 52}, 4042 (1995)
  [hep-lat/9501007].  
 %
%
\bibitem{Luscher:1998pqa} 
  M.~Luscher,
  Phys.\ Lett.\ B {\bf 428}, 342 (1998)
  [hep-lat/9802011].
 %
 %
\bibitem{Hasenfratz:1998ri} 
  P.~Hasenfratz, V.~Laliena and F.~Niedermayer,
  Phys.\ Lett.\ B {\bf 427}, 125 (1998)
  [hep-lat/9801021].
%
%
\bibitem{Neuberger:1997fp} 
  H.~Neuberger,
  Phys.\ Lett.\ B {\bf 417}, 141 (1998)
  [hep-lat/9707022].
%
%
\bibitem{Kikukawa:1998pd} 
  Y.~Kikukawa and A.~Yamada,
  Phys.\ Lett.\ B {\bf 448}, 265 (1999)
  [hep-lat/9806013];
  D.~H.~Adams,
  Annals Phys.\  {\bf 296}, 131 (2002)
  [hep-lat/9812003];
  K.~Fujikawa,
  Nucl.\ Phys.\ B {\bf 546}, 480 (1999)
  [hep-th/9811235];
  H.~Suzuki,
  Prog.\ Theor.\ Phys.\  {\bf 102}, 141 (1999)
  [hep-th/9812019].
%
%
\bibitem{Kikukawa:1998py} 
  Y.~Kikukawa and A.~Yamada,
  Nucl.\ Phys.\ B {\bf 547}, 413 (1999)
  [hep-lat/9808026].
%
\bibitem{Yang:2014sea} 
  Y.~B.~Yang, Y.~Chen, A.~Alexandru, S.~J.~Dong, T.~Draper, M.~Gong, F.~X.~Lee and A.~Li {\it et al.},
  arXiv:1410.3343 [hep-lat].
%
\bibitem{Gong:2013vja} 
  M.~Gong {\it et al.}  [XQCD Collaboration],
  Phys.\ Rev.\ D {\bf 88}, no. 1, 014503 (2013)
  [arXiv:1304.1194 [hep-ph]].
  %
  %
  \bibitem{Franz:2000ee} 
  M.~Franz, M.~V.~Polyakov and K.~Goeke,
  Phys.\ Rev.\ D {\bf 62}, 074024 (2000)
  [hep-ph/0002240].
%
%
\bibitem{Liu:1995kb} 
  K.~F.~Liu,
  hep-lat/9510046; K.~F.~Liu, S.~J.~Dong, T.~Draper and W.~Wilcox,
  Phys.\ Rev.\ Lett.\  {\bf 74}, 2172 (1995)
  [hep-lat/9406007].
    
  \end{thebibliography}
\end{document}